\documentclass[conference]{IEEEtran}

\usepackage{amsfonts}
\usepackage{setspace}

\usepackage{setspace}
\usepackage{amssymb}
\usepackage{epsfig}
\usepackage{graphicx}
\usepackage{xcolor}
\usepackage{algorithm}
\usepackage{algorithmicx}
\usepackage{algpseudocode}
\usepackage{amsmath}
\usepackage{cite}
\usepackage{subcaption}

\usepackage[normalem]{ulem}

\floatname{algorithm}{Algorithm}

\usepackage[ps2pdf,bookmarks=false,
bookmarksnumbered=false, 
bookmarksopen=false,
colorlinks=false]{}

\usepackage[utf8]{inputenc}
\usepackage[english]{babel}
\usepackage{amsthm}
\theoremstyle{definition}

\usepackage{tikz}
\newsavebox{\tempbox}

\usepackage{multicol}

\IEEEoverridecommandlockouts
\usepackage{caption,float}
\usepackage[vmargin=1cm,hmargin=1.5cm]{geometry}

\begin{document}

\title{Navigating Connected Car Cybersecurity: Location Anomaly Detection with RAN Data}
\author{Feng Wang$^*$\thanks{* Work done in the duration of internship at AT\&T Laboratories, Inc.}, Yaron Koral, Kenichi Futamura \\
AT\&T Laboratories, Inc., Middletown, NJ \\
E-mail: fwang26@syr.edu, yk216h@att.com, kf5216@att.com}
\maketitle

\begin{abstract}
The cybersecurity of connected cars, integral to the broader Internet of Things (IoT) landscape, has become of paramount concern. Cyber-attacks, including hijacking and spoofing, pose significant threats to these technological advancements, potentially leading to unauthorized control over vehicular networks or creating deceptive identities. Given the difficulty of deploying comprehensive defensive logic across all vehicles, this paper presents a novel approach for identifying potential attacks through Radio Access Network (RAN) event monitoring. The major contribution of this paper is a location anomaly detection module that identifies aberrant devices that appear in multiple locations simultaneously - a potential indicator of a hijacking attack. We demonstrate how RAN-event based location anomaly detection is effective in combating malicious activity targeting connected cars. Using RAN data generated by tens of millions of connected cars, we developed a fast and efficient method for identifying potential malicious or rogue devices. The implications of this research are far-reaching. By increasing the security of connected cars, we can enhance the safety of users, provide robust defenses for the automotive industry, and improve overall cybersecurity practices for IoT devices. 
\end{abstract}


\section{Introduction}
In 2021, the United States was home to 84 million connected cars, a figure projected to surpass 300 million by 2035 \cite{coppola2016connected, gharavi2007scanning, carfleet2022region}, making connected cars one of the most widely used cellular IoT devices. Connected cars leverage network connectivity to provide various services, from entertainment and remote unlocking to vehicle location tracking. More importantly, these networks also support critical services such as security and telematics, which control certain vehicle systems. The critical nature of these services makes connected cars enticing targets for hackers. Attacks can range from implanting malware and mimicking control channels to hijacking the vehicle's network identity, leading to severe consequences such as loss of vehicle control or denial of critical services \cite{carfrontier, carhacker, carhackjeep}.

Given the severity of these potential attacks, timely identification of compromised devices is crucial. However, this task is becoming increasingly complex due to the emergence of stealthy attack vectors \cite{becsi2015security}. Traditional signature-based detection is proven to be inadequate for connected cars as their network modules cannot run malware detection software. 

In this paper, we propose a different approach - anomaly detection through network events, which offers a rich landscape for monitoring, including registration messages, signal quality reports, and data usage. This paper primarily focuses on location anomalies, which occur when a device appears in two or more distant sites within an unrealistic travel time. For instance, if a vehicle connects to a site in Texas and then, within seconds, connects to another site in Florida, it is likely that the vehicle's network identifier has been spoofed or hijacked.
To the best of our knowledge, this is the first research endeavor that leverages network data to identify location anomalies in connected cars and detect potential security threats. We also introduce a scalable pre-filtering method for identifying abnormal devices across millions of devices, significantly reducing processing time. This research pioneers a proactive approach for fortifying the cyber-security of connected cars, significantly enhancing the safety and reliability of these widely used IoT devices.

\section{Related Work} \label{sec:relatedwork}
Diro et al. \cite{heyne2021MLanomalydetection} discuss the considerable challenges of anomaly detection for IoT devices, given their heterogeneous and distributed nature and limited resources. They warn that amid these challenges and market competition, security could be overlooked. Therefore, they argue for the necessity of monitoring and detecting anomalies in activated IoT devices and advocate for lightweight anomaly detection systems. However, their proposed machine learning techniques for anomaly detection are based on assumptions that may not be applicable in real-world scenarios, such as privacy constraints. Hence, their work lacks a clear, comparative solution that we can use as a benchmark. This research highlights the need for our paper, which focuses specifically on location-based anomalies for connected cars.

Sun et al. \cite{Sun04mobility-basedanomaly} presented an innovative methodology for detecting mobility pattern anomalies in mobile cellular users, using a high order Markov chain, a trie, and an exponentially weighted moving average (EWMA). However, their assumption that user locations are deterministically observable through Home Location Register (HLR) records is a significant limitation. In a subsequent paper \cite{Sun06mobility-basedanomaly}, they extended their work but maintained the same underlying assumption. Connected cars, as we shall discuss in Section \ref{subsec:exp}, do not conform to this assumption, making their approach less applicable to our research.

Savic et al. \cite{SavicLDBBMVSJ21} and Buthpitiya et al. \cite{Senaka_Pervasive2021} both propose machine-learning-based anomaly detection systems. Savic et al. assume that devices are GPS enabled and have uninterrupted connectivity. Buthpitiya et al. use geo-labels obtained by clustering raw GPS coordinates. However, vehicle GPS information is not included in RAN data and not available to cellular providers.

Some related works focus on location anomaly detection in traffic such as \cite{wang2019location, boquet2020variational}. The aim of \cite{wang2019location} differs from our problem as they focus on the scenario where nearby vehicles communicate location with each other, while \cite{boquet2020variational} focuses on a different anomaly detection problem. These Machine-learning-based location anomaly detection consider realistic data in a small area with heavy traffic or simulated data, with sufficient records in each location to train a machine learning model. In contrast, we consider millions of antennas and tens of millions of cars across the United States, which requires training an infeasibly large neural network with trillions of parameters. In addition,  majority of cell-sites do not experience sufficient number of connections to feed a machine learning model. Finally, machine-learning-based models lack explainability and require more computing power and longer runtimes, and thus are less welcomed by vendors. Other analytical detection methods that are viable for location anomaly detection requires timely and accurate location information, such as GPS or ad hoc networks \cite{yan2009providing}. However, connected cars usually utilize radio access network, and GPS information is not available to carriers. Therefore, in this paper, we consider the practical constraint that only estimation via cell connectivity is available.

Collectively, these studies underscore the critical need for effective anomaly detection in IoT devices, particularly connected cars. However, the assumptions and limitations inherent in these studies highlight the necessity for novel research that can tackle these challenges head-on. Our work aims to fill this gap by focusing on location-specific anomalies in connected cars, offering a fresh perspective on this pressing issue.

\section{Background} \label{sec:background}
Broadband service providers typically deploy carrier-grade probing systems to monitor, test, and manage their networks and services. These systems, designed to meet high standards of reliability, scalability, and efficiency, circumvent the need to deploy logic-based software over devices—a task that proves more challenging for IoT devices and particularly for connected cars.
Some IoT devices, including connected cars, lack Global Positioning System (GPS) capabilities, and moreover, this information is not available to carriers. Therefore, the need arises for an alternative method to identify anomalies by monitoring the locations of connected cars. This alternative involves estimating device locations based on their connection to cell tower locations. It is important to note that using cell tower location as a measure is significantly less accurate than GPS location. Additionally, connected devices may often be in an idle state, and in some instances, connect with distant cell towers due to topology or weather conditions. These behaviors result in a level of unknown mobility and variations in their Access Point Name (APN)—the name of a gateway between a device and the data network or internet.
    
Regarding security, we focus on identifiers like International Mobile Equipment Identity (IMEI) and International Mobile Subscriber Identity (IMSI). Both can be compromised by various session identifier hijacking attacks. These techniques include obtaining the parameters via an IMSI catcher \cite{IMSI_CATCHER}, using software manipulations to alter the IMEI \cite{SW_MAN}, as well as malware that extracts these parameters from the device itself \cite{burguera2011crowdroid}. Obtaining and spoofing these identifiers can be used for several attacks, such as gain illicit control of a system \cite{hwang2022web, baitha2018session,checkoway2011comprehensive}, location tracking, data theft and  denial of service attacks \cite{mavoungou2016survey}.
In this paper, we specifically focus on IMSI hijacking. This type of attack can have severe consequences, including intercepting Short Messaging Service (SMS) messages to unlock cars, detaching devices from the network, and spoofing crafted messages to cars. By addressing these challenges, our research aims to significantly enhance the security and reliability of connected car systems, thereby contributing to the broader field of IoT device security.

\section{Location Anomaly Detection Module}\label{sec:loc_anomaly}

In this section, we introduce a location anomaly detection module specifically designed for connected cars using RAN data, which are signals emitted by devices when they connect to different parts of a cellular network.
To detect a location anomaly, we examine events from each device, which we identify by its IMSI (although IMEI could serve as a viable alternative).

For each IMSI, we organize its events in chronological order, creating an approximate trajectory and analyze each pair of consecutive events for sudden, inexplicable changes in a device's location that could indicate an anomalous event.
In the remainder of this section, we will cover several key aspects of our research, introduce a scalable filtering algorithm designed for real-time anomaly detection across a large population, and discuss how to distinguish between location anomalies and non-adversarial outliers.

\subsection{RAN Event Features} \label{subsec:RAN}
 The location anomaly detection module utilizes access and mobility network records collected from communication between the device and the cell tower. These records provide valuable insights about the device's activity, revealing whether the device is in active or idle mode, and if it changes cell towers due to movement.
To estimate the device's location, we use the longitude and latitude of each cell tower. This information, extracted from the network records, forms the key features of our model, as shown in Table \ref{tab:RAN}. 
The model then examines consecutive events of the same IMSI in chronological order, verifying the feasibility of the transition between the cell towers, considering distance, speed, and handover features. Handover refers to the process by which a mobile device switches from one cell tower to another while in motion.

\begin{table}
    \small
	\begin{center}
		\begin{tabular}{ | p{28mm} || p{52mm} | }
			\hline
			Feature Name & Description  \\ \hline\hline
			record\_timestamp & Event collection timestamp   \\ \hline
			IMEI & International Mobile Equipment Identity of User Equipment (UE)  \\ \hline
			IMSI & International Mobile Subscriber Identity  \\ \hline
			current\_cell\_ID & Connected cell ID  \\ \hline
			target\_cell\_ID & Target cell ID in successful handover   \\ \hline
			source\_cell\_ID & Source cell ID in handover cases  \\ \hline
			RAN\_start\_collection \_trigger & Event triggering information collection  \\ \hline
			HO\_fail\_code & Failure reason code of last handover  \\ \hline
			first\_RTD & The first calculated Real Time Difference (RTD) value
   \\ \hline
			last\_RTD & The last calculated RTD value   \\ \hline
		\end{tabular}
		\caption{Selected RAN event features }
		\label{tab:RAN}
	\end{center}
\end{table}

\subsection{Trajectory Hash Table} \label{subsec:traj}

The first step in building the model involves creating daily trajectories for each vehicle. To achieve this, we create a list of events for each vehicle, sorting them in chronological order.
Since we analyze several billion records each day, we need a run-time-efficient data structure, to analyze the sorted trajectory list efficiently.

\begin{figure*}
  \centering
  \includegraphics[width=1\textwidth]{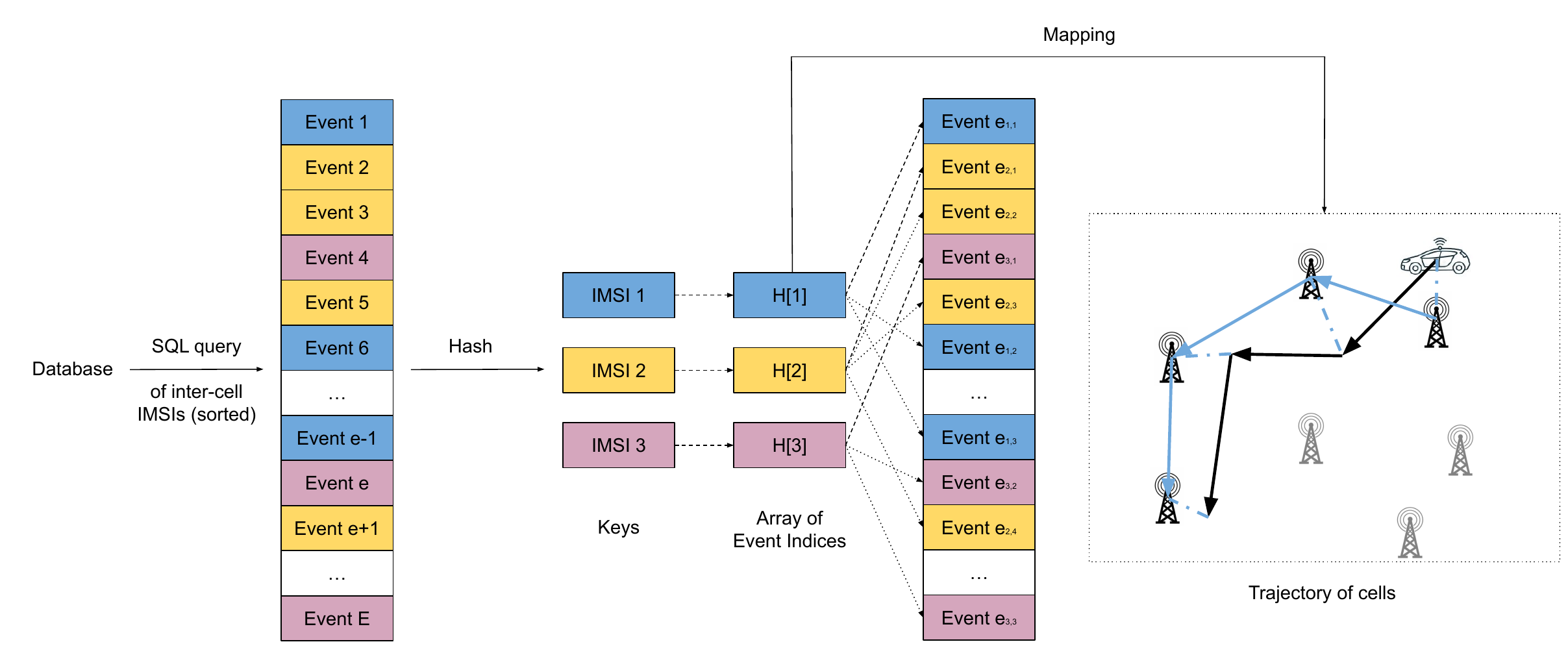}
  \caption{Diagram of establishment and usage of the trajectory Hash table.}
  \label{fig:HashTable}
\end{figure*}

We build a hash table (Fig. \ref{fig:HashTable}), denoted as $\mathcal{H}$, by iterating over the $E$ events. The key of $\mathcal{H}$ is the IMSI $i$. The value $\mathcal{H}[i] = {e_{i,1}, e_{i,2}, \ldots, e_{i,k}, \ldots, e_{i,K_i} }$ is a sequence of event indices $e \le E$ corresponding to IMSI $i$. These indices are sorted in chronological order, where $K_i$ is the number of events for IMSI $i$.
In simpler terms, for any $i$, $k_1$ and $k_2$ where $1 \le k_1 < k_2 \le K_i$, the corresponding indices $e_{i,k_1}$ and $e_{i,k_2}$ satisfy $t[e_{i,k_1}] \le t[e_{i,k_2}]$, where $t$ denotes the record\_timestamp.

Following this, we define a transition as the switch from event $e_{i,(k-1)}$ to event $e_{i,k}$. In this way, we can track the movement of each vehicle based on the cell towers it connects to, thereby allowing us to identify any anomalies in these movements.

\begin{multicols}{2}
\end{multicols}
\begin{multicols}{2}
\end{multicols}


\subsection{Location Anomaly Detection} \label{subsec:detect}
\begin{figure}
	\centering
	\includegraphics[width=1\linewidth]{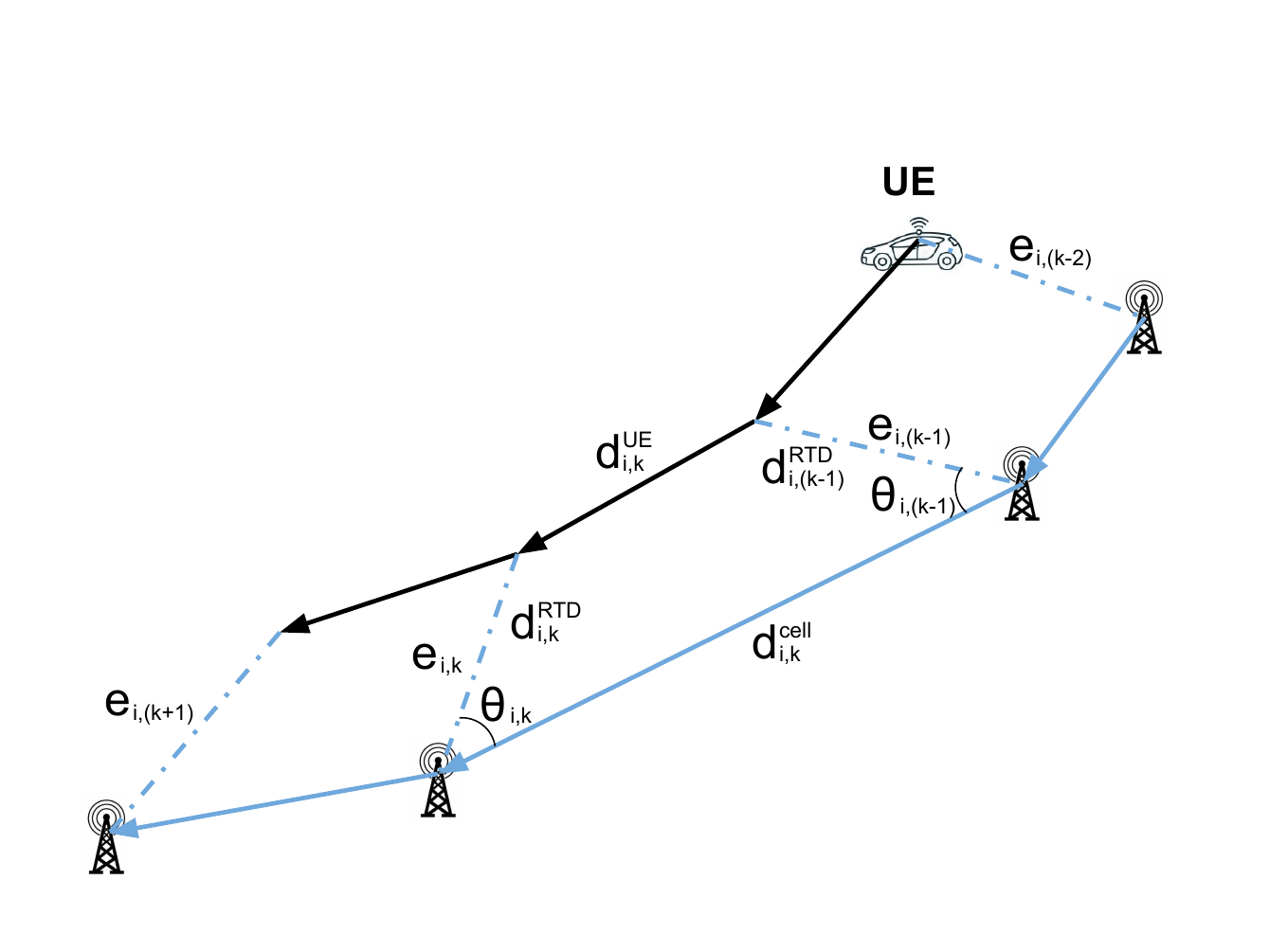}
	\caption{Distance measurement and speed estimation for location anomaly criteria. }
	\label{fig:AnomalyCriterion}
\end{figure}

A transition between two cell towers is defined as two consecutive events, $e_{i,(k-1)}$ and $e_{i,k}$, that are associated with the same IMSI $i$ and relate to two different cell tower IDs. If the estimated speed of the User Equipment (UE) exceeds a predefined physical constraint $v_{\text{max}}$ (for example, 160 km/h), we define this transition as a location anomaly. We estimate the UE speed by measuring the geographical distance between the two cell towers, the elapsed time between the two events, and the approximate UE to cell tower distance (Fig. \ref{fig:AnomalyCriterion}). We consider the distance $d^{\text{cell}}_{i,k}$ between the cell towers and the estimated lower bound of the UE speed $\hat{v}_{i,k}$, which is based on the estimated distance $\hat{d}^{\text{UE}}_{i,k}$ between the UE locations at event $e_{i,(k-1)}$ and event $e_{i,k}$. Furthermore, we use handover (HO) events to validate legitimate transitions between cell towers. 

We use  the haversine formula \cite{robusto1957cosine} to calulate the distance $d$ between two points via their latitudes and longitudes $[\text{lat}_1, \text{lon}_1]$ and $[\text{lat}_2, \text{lon}_2]$: 
\begin{equation} \label{eq:haversine1}
\Delta\phi = \frac{\pi (\text{lat}_1 - \text{lat}_2)}{180} , 
\end{equation}
\begin{equation} \label{eq:haversine2}
\Delta\lambda = \frac{\pi (\text{lon}_1 - \text{lon}_2)}{180} , 
\end{equation}
\begin{equation} \label{eq:haversine3}
a = \sin^2(\Delta\phi / 2) + \cos(\frac{\pi \text{lat}_1 }{180}) \cos(\frac{\pi \text{lat}_2 }{180}) \sin^2(\Delta\lambda / 2) , 
\end{equation}
\begin{equation} \label{eq:haversine4}
d = 2 R \text{atan2}(\sqrt{a}, \sqrt{1-a}) , 
\end{equation}
where $R=6371$ km is the earth's radius, and atan2 is the 2-argument arctangent. Therefore we have the distance $d^{\text{cell}}_{i,k}$ between cell towers of event $e_{i,(k-1)}$ and event $e_{i,k}$.

To estimate the distance from UE to cell tower we use the Real Time Difference (RTD) \cite{fischer2014observed} feature, which measures the signaling delay  between the UE and the cell tower assuming good Radio Frequency (RF) conditions and components meeting their budgeted delay error. However, the delay error can vary significantly in bad conditions, such as in mountainous terrains, where Reference Signal Received Power (RSRP) and Reference Signal Received Quality (RSRQ) -- which correlate with RTD for UE location estimation -- may be missing in the RAN event records. Therefore, to compensate for this delay error, we propose an updating queue for each cell ID, where we record the RTD compensation of the most recent $n$ events in a queue for every tower with a tolerance for up to $m$ anomalies. Whenever a new event $e_{i,k}$ is presented, the new RTD compensation will be pushed to the cell ID queue with value $C^{\text{in}}$:
\begin{equation} \label{eq:RTDcompensation}
C^{\text{in}} = d^{\text{cell}}_{i,k} - d^{\text{UEmax}}_{i,k} - d^{\text{RTD}}_{i,k} \cos{\theta_{i,k}} - d^{\text{RTD}}_{i,(k-1)} \cos{\theta_{i,(k-1)}} ,
\end{equation}
\begin{equation} \label{eq:dUEk}
d^{\text{UEmax}}_{i,k} = v_{\text{max}} ( t[e_{i,k}] - t[e_{i,(k-1)}] ) ,
\end{equation}
\begin{equation} \label{eq:dRTDk}
d^{\text{RTD}}_{i,k} = \text{RTD}_{i,k} T_s c ,
\end{equation}
where $d^{\text{cell}}_{i,k}$ is the distance between cell tower for event $e_{i,(k-1)}$ and event $e_{i,k}$. $d^{\text{UEmax}}_{i,k}$ is the estimated upper bound of the UE movement during the time, $d^{\text{RTD}}_{i,k}$ is the estimation of the distance between the cell tower and the UE at event $e_{i,k}$ based on RTD, $\theta_{i,k}$ is the angle at the cell between the direction of the antenna azimuth angle and the direction of the other cell, $v_{\text{max}}$ is the upper bound of the car speed, $\text{RTD}_{i,k}$ is the maximum of first\_RTD and last\_RTD at event $e_{i,k}$, $T_s = 1 / (15000 \times 2048)$ s is the basic time unit in Long Term Evolution (LTE), and $c$ is the speed of light. 

To test if the incoming event $e_{i,k}$ is anomalous, the $(m+1)$th highest value in the queue before the push is fed back as the RTD compensation $C^{\text{out}}_{i,k}$. Therefore, the estimated upper bound of UE speed $\hat{v}_{i,k}$ is
\begin{equation} \label{eq:vik}
\footnotesize
\hat{v}_{i,k} = \frac{d^{\text{cell}}_{i,k} - d^{\text{RTD}}_{i,k} \cos{\theta_{i,k}} - d^{\text{RTD}}_{i,(k-1)} \cos{\theta_{i,(k-1)}} - ( C^{\text{out}}_{i,k} + C^{\text{out}}_{i,k-1}) / 2}{ t[e_{i,k}] - t[e_{i,(k-1)}] } .
\end{equation}
Furthermore, valid HO transitions are excluded from potential location anomaly cases even if the distance is large.

To summarize, we judge the transition from event $e_{i,(k-1)}$ to event $e_{i,k}$ as location anomaly if it satisfies the following criteria:
\begin{equation} \label{eq:criterion}
\left(d^{\text{cell}}_{i,k} > d_{\text{min}}\right) \land \left(\hat{v}_{i,k} > v_{\text{max}}\right) \land \left( \neg O^{\text{HFC}}\right) \land \left( \neg O^{\text{SCT}}\right) \land \left( \neg O^{\text{TGT}}\right) ,
\end{equation}
where $O^{\text{HFC}}$ is the statement that HO\_fail\_code indicates one of the HO types, $O^{\text{SCT}}$ is the statement that RAN\_start\_collection\_trigger involves HO, $O^{\text{TGT}}$ is the statement that the current\_cell\_ID at event $e_{i,k}$ is the same as target\_cell\_ID at event $e_{i,(k-1)}$, and $d_{\text{min}}$ and $v_{\text{max}}$ are adjustable parameters depending on the site types and morphology (urban, suburban, or rural area) of both cells. That is to say, a non-handover transition with sufficiently long distance and sufficiently fast speed is considered a location anomaly.

\subsection{Filtering Algorithm for Scalability} \label{subsec:algorithm}
While equation (\ref{eq:criterion}) provides a reliable method for anomaly identification, the volume of RAN events makes applying this across tens of millions of connected car devices and billions of events impractical. Therefore, we propose a filtering algorithm based on the Non-access stratum (NAS) events between the device and the core network, which significantly reduces processing time. 

Our proposed scalable algorithm is shown in Algorithm \ref{alg:filter}. First, we filter the full set of connected cars with NAS events to $\mathcal{I}^{\text{all}}$, so only the IMSIs that were active and moved across a certain distance are selected. Second, we detect potential location anomalies using only cell locations and time stamps of involving NAS events, and keep a subset $\mathcal{I}^{\text{NAS}} \subseteq \mathcal{I}^{\text{all}}$. We note that this set of criteria are much looser than that in equation (\ref{eq:criterion}) as the number of NAS events is less than all RAN events, and RAN features including RTD are ignored. By itself, this pre-filter could result in a high false positive rate, i.e., the filtered set of IMSIs $\mathcal{I}^{\text{NAS}}$ may not necessarily be location anomalies. However, by avoiding many time-consuming tests and concentrating on certain critical events, we significantly narrow down the search range and reduce processing time, i.e., $|\mathcal{I}^{\text{NAS}}| \ll |\mathcal{I}^{\text{all}}|$. Finally, we perform the aforementioned location anomaly detection as in (\ref{eq:criterion}) on all RAN events with all features of the selected set $\mathcal{I}^{\text{NAS}}$ of IMSIs, and get the set $\mathcal{I}$ of IMSIs with location anomalies. 

\begin{algorithm}[h]
\caption{Location anomaly detection algorithm based on RAN events, pre-filtered via NAS events for scalability}
	\label{alg:filter}
	\begin{algorithmic}[1]
	    \State{Extract $\mathcal{I}^{\text{all}}$, the set of all distinct IMSI $i$ of IoT cars with inter-cell NAS events in the day.}
	    
	    \State{Consider only cell ID and time stamp of \textbf{NAS events}, and sort trajectories $\forall i \in \mathcal{I}^{\text{all}}, \mathcal{H}[i] = \{e_{i,1}, e_{i,2}, \ldots, e_{i,k}, \ldots, e_{i,K_i} \} $.}
	    \State{Extract $\mathcal{I}^{\text{NAS}} = \{ i \in \mathcal{I}^{\text{all}} \ | \ \exists k, (d^{\text{cell}}_{i,k} > d_{\text{min}}) \land (\frac{d^{\text{cell}}_{i,k}}{ t[e_{i,k}] - t[e_{i,(k-1)}] } > v_{\text{max}})  \} $.}
	    
	    \State{Consider \textbf{all RAN events}, and sort trajectories $\forall i \in \mathcal{I}^{\text{NAS}}, \mathcal{H}[i] = \{e_{i,1}, e_{i,2}, \ldots, e_{i,k}, \ldots, e_{i,K_i} \} $.}
	    \State{Extract $\mathcal{I} = \{ i \in \mathcal{I}^{\text{NAS}}  \ | \  \exists k, (d^{\text{cell}}_{i,k} > d_{\text{min}}) \land (\hat{v}_{i,k} > v_{\text{max}}) \land ( \neg O^{\text{HFC}}) \land ( \neg O^{\text{SCT}}) \land ( \neg O^{\text{TGT}})  \} $.}
	    
	    \State{\textbf{Return} $\mathcal{I}$. }
	\end{algorithmic}
\end{algorithm}

While the aforementioned pre-filtering algorithm reduces the runtime on daily data from years to days, it still does not meet our detection delay requirement. Therefore, we implement the trajectory and cell location tables as Hash tables for Algorithm \ref{alg:filter}, reducing index look up from $O(n)$ to  $O(1)$ time complexity. With these improvements, the algorithm detects anomalies across tens of millions of connected cars within half an hour. 

\subsection{Corner Cases Analysis} \label{subsec:exp}
 
\begin{figure}
	\centering
	\includegraphics[width=1\linewidth]{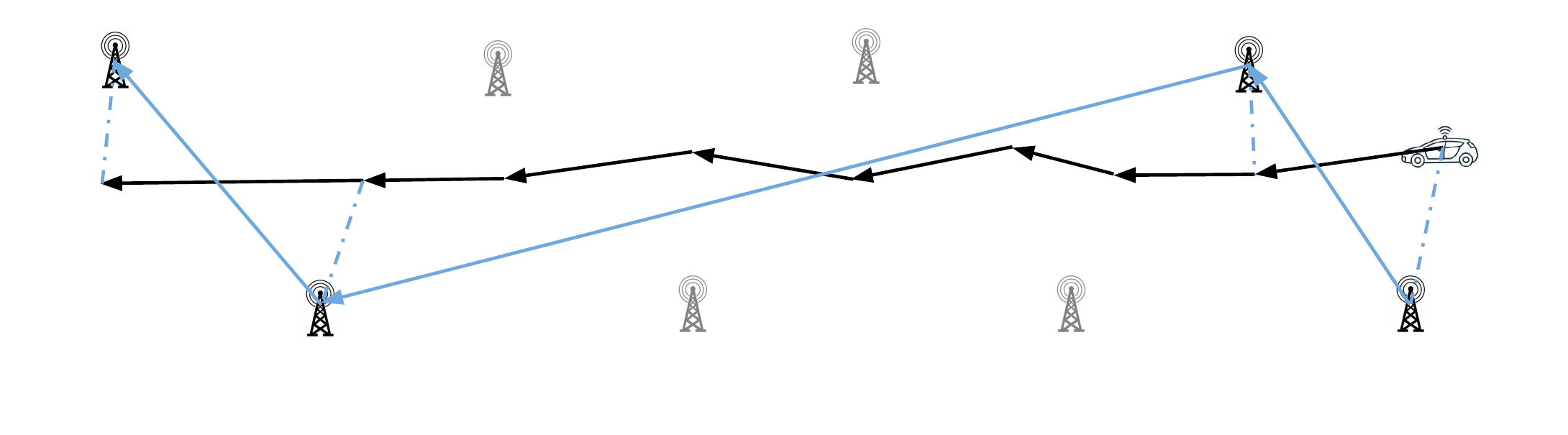}
	\caption{IoT cars are usually in idle state: large gap between event 2 and event 3. }
	\label{fig:AbnormalCase0}
\end{figure}

Our model takes into account a few corner cases that arise from abnormal radio conditions and the unique connectivity patterns of vehicle modems. A key consideration is that a vehicle modem can remain idle for extended periods, during which it may pass through several cell towers without connecting to them.
Despite the expectation that most valid transitions between cell towers would involve a HO event, about half of the transitions in both metropolitan and rural areas do not involve an HO event. In fact about 90\% of the RAN events are re-connections rather than HOs. As illustrated in Fig. \ref{fig:AbnormalCase0}, there are proper HOs between events 1 and 2, and between events 3 and 4. However, there is no HO between events 2 and 3. This is because the UE becomes idle during that time, and event 3 merely represents a Radio Resource Control (RRC) re-establishment. 

\begin{figure}
	\centering
	\includegraphics[width=0.6\linewidth]{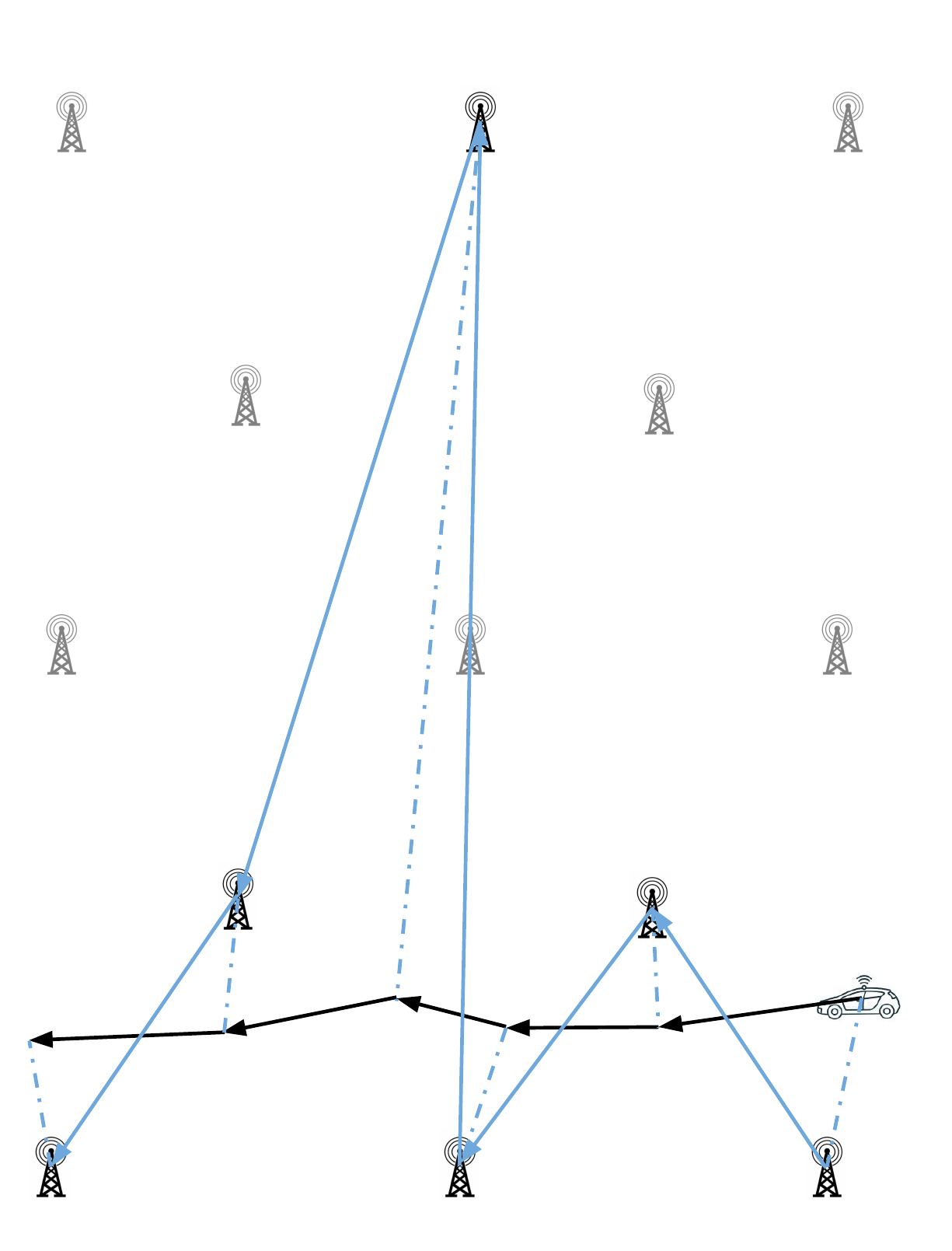}
	\caption{Sudden switching to a distant cell and switching back due to heavy traffic, reflective terrain, or rolling terrain. }
	\label{fig:AbnormalCase1}
\end{figure}

Another challenge arises when a device connects to a distant cell tower due to heavy traffic or challenging terrain, such as rivers, lakes, or mountains. In Fig. \ref{fig:AbnormalCase1}, the trajectory switches to a cell far away before switching back to a nearby cell. While the event switching away may be HO, switching back is usually via re-connection. To address this, we calibrate our estimation of the UE's location using RTD. The distance estimation via RTD is more accurate across plain and water surface, but is usually underestimated in mountainous area. In the latter case, we compensate for the distance estimation via $C^{\text{out}}_{i,k}$ from the aforementioned queue for the specific cell. 

The RTD may vary even for the same UE at similar locations and times. This is particularly noticeable in cases where a UE connects to the same cell ID in two consecutive events, resulting in a significant change in the estimated distance within a few seconds. As described in our algorithm, we made an assumption that for any $n$ consecutive events of any cell ID, there is a high confidence that there are no more than $m$ location anomalies. Thus, we use the $(m+1)${th} highest value in the queue as the new compensation $C^{\text{out}}_{i,k}$. At the beginning of data collection for the queue, one may uniformly initialize with $C=20$km.

By addressing these corner cases, our model offers a more robust and accurate approach to detecting location anomalies in connected cars.

\section{Experimental Results} \label{sec:experimental}

In accordance with our commitment to privacy and security, certain types of information cannot be disclosed in this paper. Consequently, statements that might imply vulnerabilities for specific brands, such as suggesting that a particular car manufacturer's vehicles are prone to identity spoofing, are not included. Our focus remains on presenting our methodology and findings without infringing upon these ethical and privacy considerations. We note that the anomaly detection system alerts on the existence of a location anomaly and id spoofing. It does not report the actual location due to privacy concerns.

In Table \ref{tab:runtime_compare}, we show the scalability of our proposed algorithm using a Java implementation, presenting overall runtime and the local peak memory usage during processing. The data is collected on 03/14/2024 across 12 vendors with 14,099,905 devices and 69,083,615 NAS events in total. The detection without filtering and the logic-based algorithm \cite{dietzel2014flexible} took over one day to process one day's worth of data and, therefore, are not practically useful.  In contrast, our proposed method processed the data within half an hour.
\begin{table}
    \small
	\begin{center}
		\begin{tabular}{ | p{10mm} || p{20mm} | p{22mm} | p{22mm} | } \hline
		 & Our detection with filtering & Our detection without filtering & Logic-based algorithm \cite{dietzel2014flexible}  \\ \hline\hline
		Runtime  & 27.95 min & $>1$ day & $>1$ day  \\ \hline
        Memory & 9 GB & - & -  \\ \hline
		\end{tabular}
    	\caption{Comparison of overall runtime and peak memory usage between different algorithms on daily data from 12 vendors. }
    	\label{tab:runtime_compare}
	\end{center}
\end{table}

In Table \ref{tab:parameter}, we provide the number of detected anomalous devices by our proposed algorithm, comparing the impacts of car speed upper bound $v_{\text{max}}$ and distance compensation $C^{\text{out}}_{i,k}$ -- we show the average $\overline{C^{\text{out}}_{i,k}} = \frac{1}{K} \sum^K_{k=1}{C^{\text{out}}_{i,k}}$ over all cell towers. The data is collected from two different vendors on 03/14/2024 across 1,235,661 devices and 13,923,702 NAS events. The distance compensation is determined by the assumed ratio of anomaly events $m/n$. While general ground truth unknown for all cars, we reviewed trajectories of thousands of cars to identify if they are anomalous devices, test devices or non-anomaly corner cases. The result of the trade-off between precision and recall is in the elbow point at $m/n = 2.5 \times 10^{-6}$ which gives us distance compensations where $\overline{C^{\text{out}}_{i,k}} = 16$ km, and $v_{\text{max}} = 160$ km/h.
\savebox{\tempbox}{\begin{tabular}{@{}r@{}l@{\space}}
& $v_{\text{max}}$ \\ $\overline{C^{\text{out}}_{i,k}}$
\end{tabular}}
\begin{table}
    \small
	\begin{center}
		\begin{tabular}{ | p{17mm} || p{15mm} | p{15mm} | p{15mm} | } \hline
		\tikz[overlay]{\draw (-0.21,\ht\tempbox) -- (1.99,-\dp\tempbox);} \usebox{\tempbox}\hspace{\dimexpr 1pt-\tabcolsep} & 80 km/h & 160 km/h & 240 km/h  \\ \hline\hline
		8 km  & 44, 2 & 2, 2 & 2, 2  \\ \hline
        16 km & 33, 2 & 2, 2 & 2, 2  \\ \hline
        24 km  & 23, 2 & 2, 2 & 2, 2  \\ \hline
		\end{tabular}
    	\caption{Number of daily anomalies with proposed algorithm from two vendors, given different speed threshold $v_{\text{max}}$ and different averaged distance compensation $\overline{C^{\text{out}}_{i,k}}$. }
    	\label{tab:parameter}
	\end{center}
\end{table}

\begin{figure}
	\centering
	\includegraphics[width=1\linewidth]{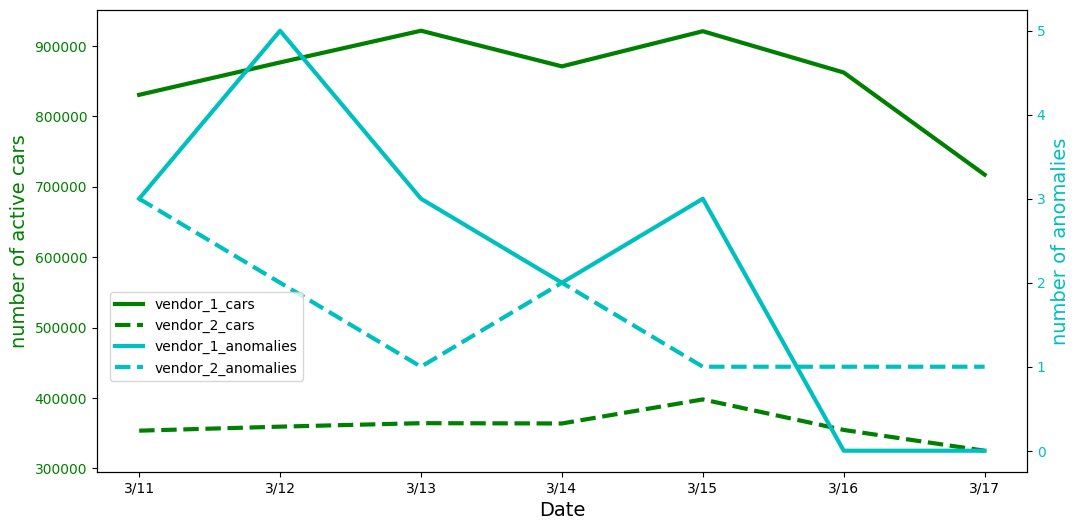}
	\caption{The number of active car devices and the number of devices with location anomaly detected for two vendors over a week.}
	\label{fig:curve_over_week}
\end{figure}

In Fig. \ref{fig:curve_over_week}, we show the number of active car devices that connected to multiple cells and the number of devices with location anomalies for two vendors during a week in 2024. Generally speaking, there are fewer activated cars and fewer anomalies during weekends compared to weekdays. However, vendors with more cars do not necessarily have more location anomalies.

The location anomaly detection algorithm analyzed three months worth of actual network data. The data included signaling RAN events that were generated by connected cars that operate over the AT\&T network. Overall, the network includes tens of millions of connected cars. 

Initially we analyzed all daily records of 10,000 cars. The process took roughly 10 days to analyze location anomalies for a single day. Using the Trajectory Hash Table from Section \ref{subsec:traj} and the Filtering Algorithm from Section \ref{subsec:algorithm}, we were able to analyze daily anomalies of 8,000,000 cars in less than half an hour.

Through our analysis, we identified tens of location anomalies that appeared to be spoofed identifier cases. A device with an IMSI that is spoofed typically uses a single IMEI but may appear to travel hundreds of kilometers within a few minutes before returning instantaneously to its original location, indicating that the device was spoofed from another city. Typically in the United States, more anomalies occurred on the west coast than the east coast and less frequently in mid-west and south-west regions.


\section{Conclusion}\label{sec:con}
This paper has introduced a novel approach to location anomaly detection for connected cars using Radio Access Network (RAN) event monitoring. We have outlined the challenges in detecting location anomalies and shown how our model addresses these issues, demonstrating the capability of our model to handle large volumes of data and complex connectivity patterns. Furthermore, we have detailed the corner cases our model takes into account, showing its robustness in the face of abnormal radio conditions and the unique connectivity patterns of vehicle modems. Despite the challenges, our research has shown that location anomaly detection is not only feasible but also highly effective for enhancing the security and reliability of connected cars. Our work serves as a pioneering step in this domain, contributing significantly to the broader field of connected car security. We encourage future machine learning based research to focus on topology and improve the location anomaly detection performance with heterogeneous traffic where the majority of locations have low traffic volume.

\bibliographystyle{IEEEtran}
\bibliography{ref.bib}

\end{document}